# The kagomé metals $RbTi_3Bi_5$ and $CsTi_3Bi_5$

Dominik Werhahn[a], Brenden R. Ortiz[b], Aurland K. Hay[b], Stephen D. Wilson[b], Ram Seshadri[b], and Dirk Johrendt[a],*

[a]Department Chemie, Ludwig-Maximilians-Universität München, Butenandtstraße 5–13, 81377 München, Germany; E-mail: johrendt@lmu.de

[b]Materials Department, Materials Research Laboratory and California Nanosystems Institute, University of California Santa Barbara, Santa Barbara, CA 93106, USA

**Abstract:** The kagomé metals $RbTi_3Bi_5$ and $CsTi_3Bi_5$ were synthesized both as polycrystalline powders by heating the elements an argon atmosphere and as single crystals grown using a self-flux method. The compounds crystallize in the hexagonal crystal system isotypically to $KV_3Sb_5$ (*P*6/*mmm*, *Z* = 1, $CsTi_3Bi_5$: *a* = 5.7873(1), *c* = 9.2062(1) Å; $RbTi_3Bi_5$: *a* = 5.773(1), *c* = 9.065(1) Å). The titanium atoms form a kagomé net with bismuth atoms in the hexagons as well as above and below the triangles. The alkali metal atoms are coordinated by 12 bismuth atoms and form $AlB_2$-like slabs between the kagomé layers. Magnetic susceptibility measurements with $CsTi_3Bi_5$ and $RbTi_3Bi_5$ single crystals reveal Pauli-paramagnetism and traces of superconductivity caused by $CsBi_2$/$RbBi_2$ impurities. Magnetotransport measurements reveal conventional Fermi liquid behavior and quantum oscillations indicative of a single dominant orbit at low temperature. DFT calculations show the characteristic metallic kagomé band structure similar to that of $CsV_3Sb_5$ with reduced band filling. A symmetry analysis of the band structure does not reveal an obvious and unique signature of a nontrivial topology.

## 1 Introduction

The kagomé net of corner-connected triangles (Figure 1) is a common structural motif in minerals [1] and intermetallic compounds like the Laves phases [2]. It was first named by the Japanese physicist Itiro Syôzi because of its similarity to the woven patterns of bamboo baskets [3]. His work already fore shadowed that kagomé nets may exhibit exceptional properties since he showed that, unlike in square and hexagonal nets, antiferromagnetism is highly frustrated within the kagomé net.

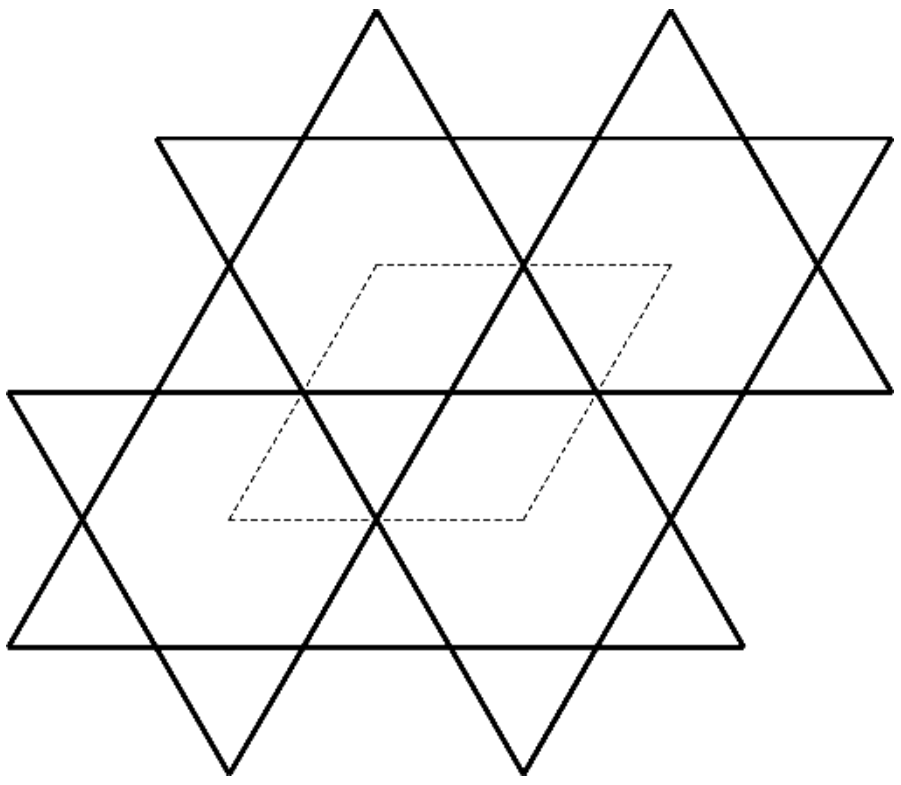

**Figure 1:** The four-connected kagomé net of corner-sharing triangles (plane group *p*6*mm*).

Later it turned out that its inherent spin frustration can indeed result in exceptional electronic or magnetic ground states. One of them is the quantum spin liquid (QSL), a theoretical concept introduced by P. W. Anderson who proposed it as an alternative to the antiferromagnetic Néel state [4]. Magnetic order in kagomé lattices can be induced by quantum fluctuations, a special case of order from disorder [5]. The theoretical concept was further developed under the motivation to explain high-$T_c$ superconductivity [6]. However, the experimental realization of the QSL turned out to be difficult, and in spite of many extensively studies [7–14], the question whether it exists or not remains under debate [15]. Another peculiarity is that itinerant (delocalized) electrons on a kagomé lattice form a Dirac semimetal like graphene [16]. Therefore kagomé lattices have attracted much attention as a base for the realization of correlated and topological quantum states like charge fractionalization [17], density waves [16] and superconductivity [18]. The recently discovered kagomé metals $A$V$_3$Sb$_5$ ($A$ = K, Rb, Cs) [19] showed particularly rich physics like unconventional chiral charge order with topological band structures [20], superconducting ground states [21, 22] as well as a giant extrinsic anomalous Hall effect [23]. It is quite interesting to examine the properties of the KV$_3$Sb$_5$-type compounds as a function of the band filling. Hole doping with tin revealed increasing superconducting $T_c$'s during rapid suppression of the charge density waves in KV$_3$Sb$_{5-x}$Sn$_x$ ($0 \leq x \leq 1$) [24] and double-dome superconductivity in CsV$_3$Sb$_{5-x}$Sn$_x$ ($0 \leq x \leq 1$) [25]. Apparently, the properties change strongly with the electron count, and although other variants of the AM$_3$X$_5$ family are predicted, only the vanadium antimonides AV$_3$Sb$_5$ are currently known [26]. During a systematic investigation we discovered RbTi$_3$Bi$_5$ and CsTi$_3$Bi$_5$ which crystallize in the KV$_3$Sb$_5$-type structure and have a significantly reduced band filling. Here we present the synthesis, single-crystal growth, crystal structure, basic magnetic and transport properties as well as the electronic band structure of the new members of the KV$_3$Sb$_5$-type family.

## 2 Results

### 2.1 Synthesis and crystal structure

Polycrystalline samples of RbTi$_3$Bi$_5$ and CsTi$_3$Bi$_5$ form by heating stoichiometric mixtures of the elements at $T$ = 600°C in alumina crucibles welded in silica tubes under an atmosphere of purified argon. The resulting black powders are extremely sensitive to air and decompose rapidly. Single crystals of RbTi$_3$Bi$_5$ and CsTi$_3$Bi$_5$ were synthesized using a self-flux method published previously [27]. Crystals form as hexagonal plates with typical dimensions of up to 10 × 10 × 0.25 mm$^3$. They are malleable, readily exfoliated and extremely sensitive to air and water. A small single crystal suitable for X-ray diffraction was selected from the CsTi$_3$Bi$_5$ powder sample and sealed in a glass capillary under argon atmosphere. The structure was solved and refined in the hexagonal space group $P6/mmm$ which resulted in the crystal structure isotypic to KV$_3$Sb$_5$. The structure of RbTi$_3$Bi$_5$ was refined from X-ray powder data using the atom coordinates of $CsTi_3Bi_5$ as starting parameters ($RbTi_3Bi_5$: $a$ = 5.773(1), $c$ = 9.065(1) Å, $z_{Bi2}$ = 0.2335(4), $R_p$ = 0.025, GoF = 0.961). The main results of the single-crystal structure determination are compiled in Tables 1–3. X-ray powder diffraction patterns of both compounds contain small amounts (~1–3 wt%) of the cubic Laves phases $RbBi_2$ and $CsBi_2$ as the only impurity, respectively (Figure 2). Due to the large X-ray absorption, the powders in the capillaries have been diluted with ground silica, which adds a diffuse background.

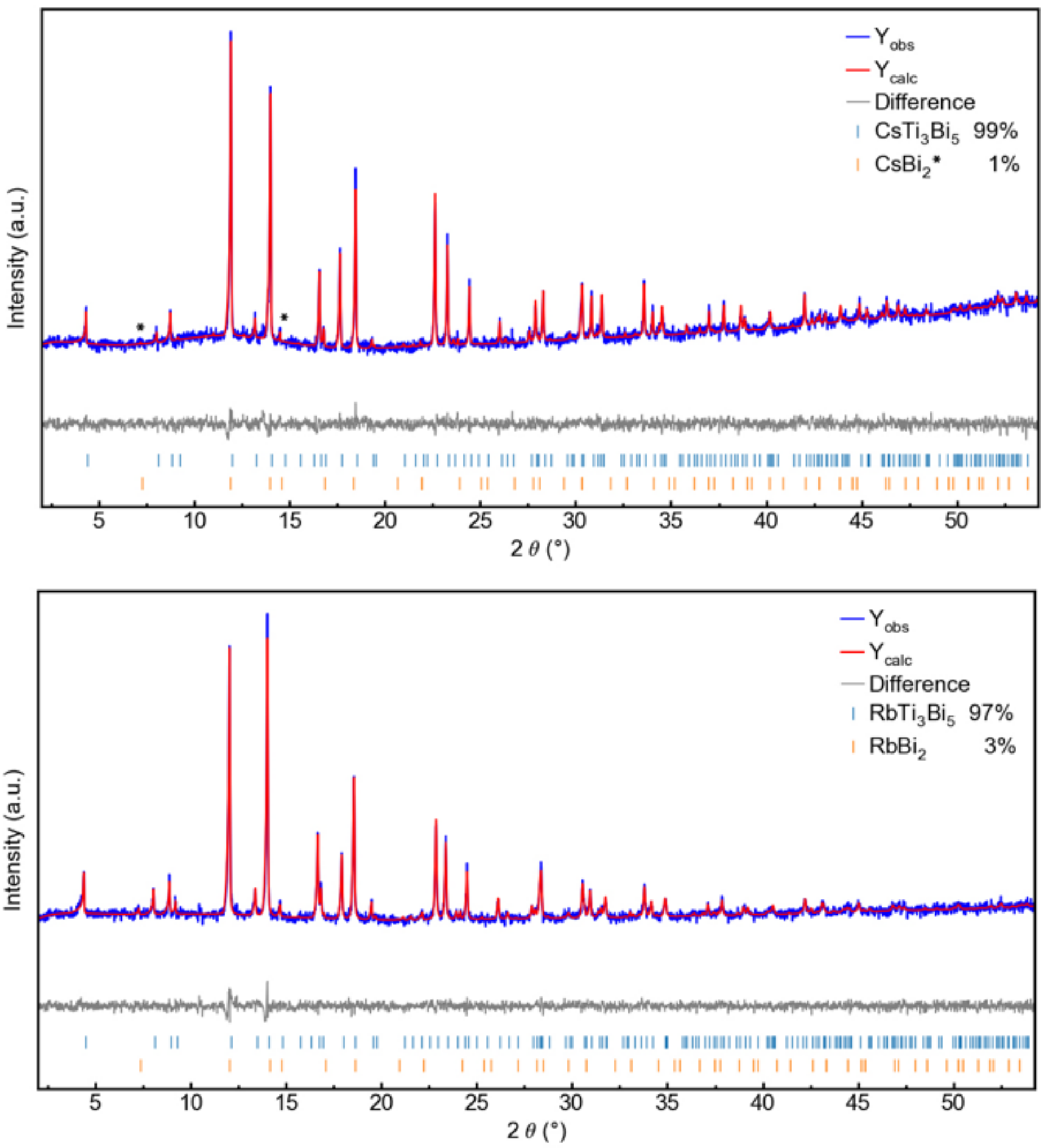

**Figure 2:** X-ray powder patterns (Mo$K\alpha_1$ radiation, blue circles) with Rietveld fits (red) and difference curves (grey) of $CsTi_3Bi_5$ (top) and $RbTi_3Bi_5$ (bottom).

**Table 1:** Crystallographic data of $CsTi_3Bi_5$

| | |
|---|---|
| Empirical Formula | $CsTi_3Bi_5$ |
| Color and habitus | black plate |
| Molecular mass, g mol$^{-1}$ | 1321.51 |
| Crystal size, mm$^3$ | 0.12 × 0.12 × 0.04 |
| Crystal system | hexagonal |
| Space group | *P*6/*mmm* (no. 191) |
| $a$, Å | 5.7873(3) |
| $c$, Å | 9.2062(6) |
| $V$, Å$^3$ | 267.03(3) |
| $Z$ | 1 |
| $\rho_{calcd}$, g cm$^{-3}$ | 8.22 |
| $\mu$(Mo$K\alpha$), cm$^{-1}$ | 87.6 |
| $F$(000), $e$ | 536 |
| $hkl$ range | $-7 \le h \le 8$, $-8 \le k \le 8$, $-13 \le h \le 13$ |
| $((\sin\theta)/\lambda)_{max}$, Å$^{-1}$ | 0.715 |
| Reflexions measured | 5093 |
| Reflexions unique | 218 |
| $R_{int}/R_{sigma}$ | 0.118 / 0.052 |
| Parameters refined | 11 |
| $R(F)^a$ ($I > 4\ \sigma(I)$ / all data) | 0.047 / 0.051 |
| $wR(F^2)^b$ (all data) | 0.106 |
| GoF $(F^2)^c$ | 1.24 |
| $\Delta\rho_{fin}$ (max/min), $e$ Å$^{-3}$ | 4.11 / −4.01 |

[a] $R(F) = \Sigma||F_o|-|F_c||/\Sigma|F_o|$; [b] $wR(F^2) = [\Sigma w(F_o^2-F_c^2)^2/\Sigma w(F_o^2)^2]^{1/2}$; $w = [\sigma^2(F_o^2)+(AP)^2+BP]^{-1}$, where $P = (\mathrm{Max}(F_o^2, 0)+2F_c^2)/3$; [c] GoF = $S = [\Sigma w(F_o^2-F_c^2)^2/(n_{obs}-n_{param})]^{1/2}$.

**Table 2:** Atom positions and displacement parameters (Å$^2$) with estimated standard deviations in parenthesis. Data of $RbTi_3Bi_5$ are from Rietveld fit of the X-ray powder diffraction pattern.

| Atom | Wyckoff | *x* | *y* | *z* | $U_{eq}$ |
|---|---|---|---|---|---|
| Rb | 1*a* | 0 | 0 | 0 | 0.027 |
| Cs | | | | | 0.028(1) |
| Ti | 3*g* | 1/2 | 1/2 | 1/2 | 0.013 |
| | | | | | 0.009(1) |
| Bi1 | 4*h* | 1/3 | 2/3 | 0.2335(4) | 0.015 |
| | | | | 0.2387(1) | 0.013(1) |
| Bi2 | 1*b* | 0 | 0 | 1/2 | 0.013 |
| | | | | | 0.011(1) |

**Table 3:** Selected bond lengths (Å) with estimated standard deviations in parenthesis.

| | | $RbTi_3Bi_5$ | $CsTi_3Bi_5$ |
|---|---|---|---|
| Rb/Cs–Bi1 | 12× | 3.948(2) | 3.9990(6) |
| Ti–Bi1 | 4× | 2.935(3) | 2.9290(9) |
| Ti–Bi2 | 2× | 2.887(1) | 2.8937(2) |
| Ti–Ti | 4× | 2.887(2) | 2.8937(2) |
| Bi1–Bi1 | 3× | 3.3333(4) | 3.3413(2) |

Figure 3 shows the crystal structure of $CsTi_3Bi_5$. The titanium atoms form a planar kagomé net ($d_{Ti–Ti}$ = 2.8937(2) Å) with bismuth atoms in the centers of the hexagons ($d_{Ti–Bi2}$ = 2.8937(2) Å, Fig. 3a) as well as above and below the triangles ($d_{Ti–Bi1}$ = 2.929(1) Å, Fig. 3b). These bismuth atoms form face-sharing hexagonal prisms around the cesium atoms ($d_{Cs–Bi1}$ = 3.999(1) Å Fig. 3c) that leads to 12-fold coordination of cesium. This motif is well known from numerous $AlB_2$-type compounds [28]. One can thus describe the structure of $CsTi_3Bi_5$ as stacking of $CsBi_4$ layers and planar $Ti_3Bi$ kagomé nets (Fig. 3c).

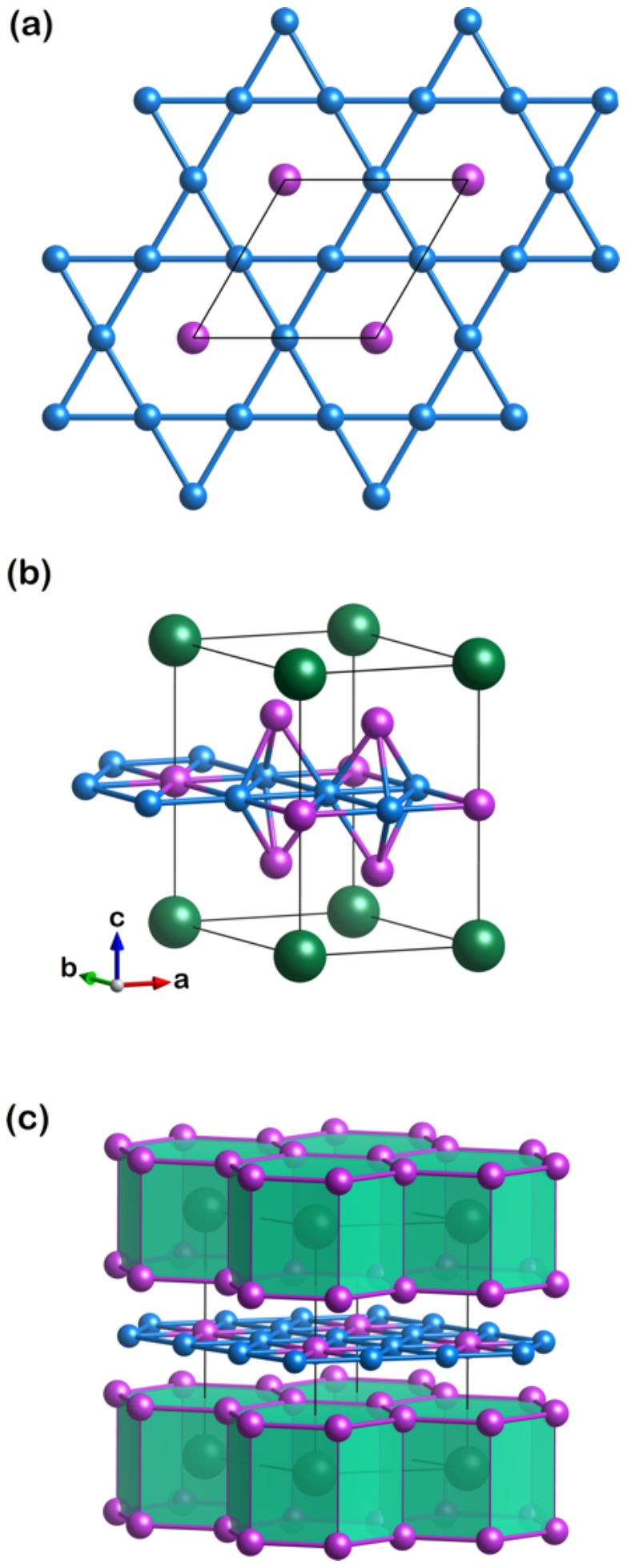

**Figure 3:** Crystal structure of $CsTi_3Bi_5$ (Cs green, Ti blue, bismuth violet). (a) Kagomé net of titanium atoms with bismuth atoms in the hexagons. (b) Unit cell with additional bismuth atoms above and below the triangles of the kagomé net. (c) Cesium atoms in face-sharing hexagonal prisms of bismuth atoms.

Recently Schoop *et al.* proposed a scheme to classify kagomé compounds based on geometric and electronic considerations [29]. A tolerance factor $t = \frac{d_K}{d_{NN}}$ is defined where $d_K$ is the bond length within the kagomé network and $d_{NN}$ is the bond length between the atoms of the kagomé network and their nearest neighbors. This gives $t = 0.893$ for $RbTi_3Bi_5$ and $t = 0.988$ for $CsTi_3Bi_5$, respectively. Values ≤ 1.1 indicate that the in-plane bonding is strong, and the band structure is expected to be dominated by the orbitals from the kagomé plane. Such a crystal features an isolated, quasi two-dimensional lattice. Thus, $RbTi_3Bi_5$ and $CsTi_3Bi_5$ fall in the same category as the vanadium compounds $AV_3Sb_5$ ($A$ = K, Rb, Cs) [19].

**2.2 Magnetic and electrical transport properties**

Figure 4 shows zero-field cooled magnetic susceptibilities collected under $\mu_0 H = 1$ Tesla external field and magnetization isotherms at $T = 1.8$ K and 300 K for $RbTi_3Bi_5$ and $CsTi_3Bi_5$ single crystals, respectively (inserts). Both compounds exhibit weak and only slightly temperature-dependent paramagnetism in the order of magnitude of $3.6 \times 10^{-4}$ emu mol$^{-1}$ Oe$^{-1}$ (1 Oe = 79.6 A m$^{-1}$) which increases at temperatures below 20 K. The susceptibilities are typical for Pauli paramagnetic metallic materials with traces of local moment paramagnetic impurities, which account for the upturn at low temperatures. The magnetization loops are accordingly linear with increasing field and show no peculiarities (inserts in Fig. 4). In Figure 4 we also plot the data for $RbV_3Sb_5$ and $CsV_3Sb_5$ for comparison, which exhibit CDW anomalies and structural phase transitions near 90 K. Such anomalies are not visible in the titanium compounds, while the susceptibilities at higher temperatures are similar.

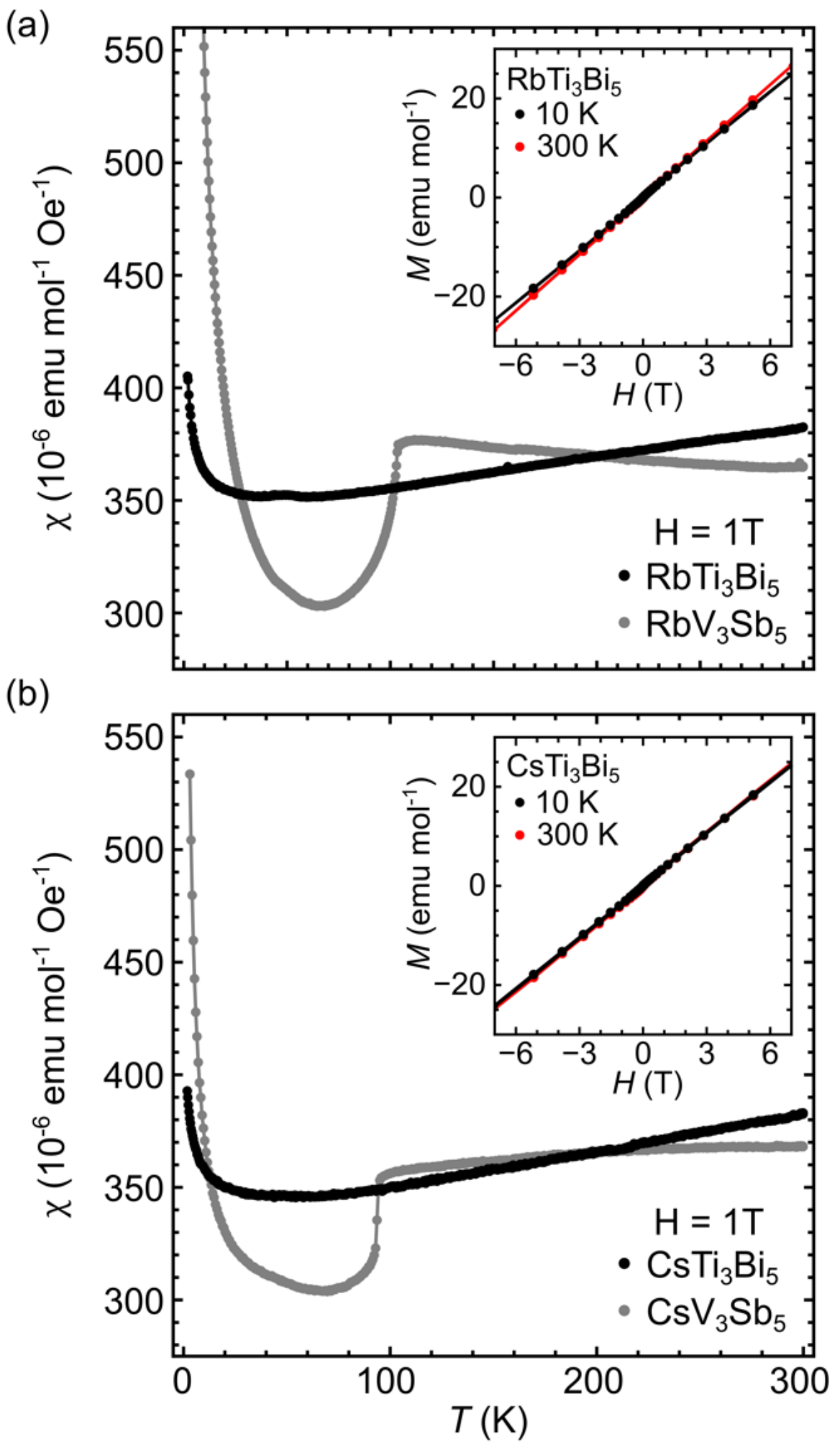

**Figure 4:** Magnetic susceptibilities of $RbTi_3Bi_5$ and $CsTi_3Bi_5$ single crystals at 1 Tesla external field. The inserts show magnetization isotherms at $T$ = 1.8 and 300 K. The data for $RbV_3Sb_5$ and $CsV_3Sb_5$ are plotted for comparison.

The electrical resistivity of the single crystals with current flowing in the *ab* plane decrease nearly linearly with temperature and saturate towards 3–5 μΩ cm at 1.8 K as expected of metals (Figure 5). Residual resistance ratios (RRR) are 12 for $RbTi_3Bi_5$ and 18 for $CsTi_3Bi_5$ crystals, respectively. Both materials show weak drops of the resistance near 4 K which reflects small inclusions of a superconducting impurity phase, and no-zero resistivity state was observed. Since we do not detect significant diamagnetism in the low-field susceptibility at 0.5 mT, the trace signatures of superconductivity most likely arise from traces of $RbBi_2$ and $CsBi_2$, respectively, incorporated in the crystals and observed as impurities in the powder samples. The binary Laves phases $RbBi_2$ and $CsBi_2$ are known to be superconducting at 4.21 and 4.65 K, respectively [30, 31], which match the temperatures of the small dips in the resistance. In order to fully rule out possible bulk superconductivity, we measured the heat capacity of powders. As seen from Figure 6, the heat capacity is featureless and saturates to the Dulong-Petit value at room temperature as expected. The Sommerfeld coefficient was extracted from the low temperature data to $\gamma = 13.9$ mJ mol$^{-1}$ K$^{-2}$.

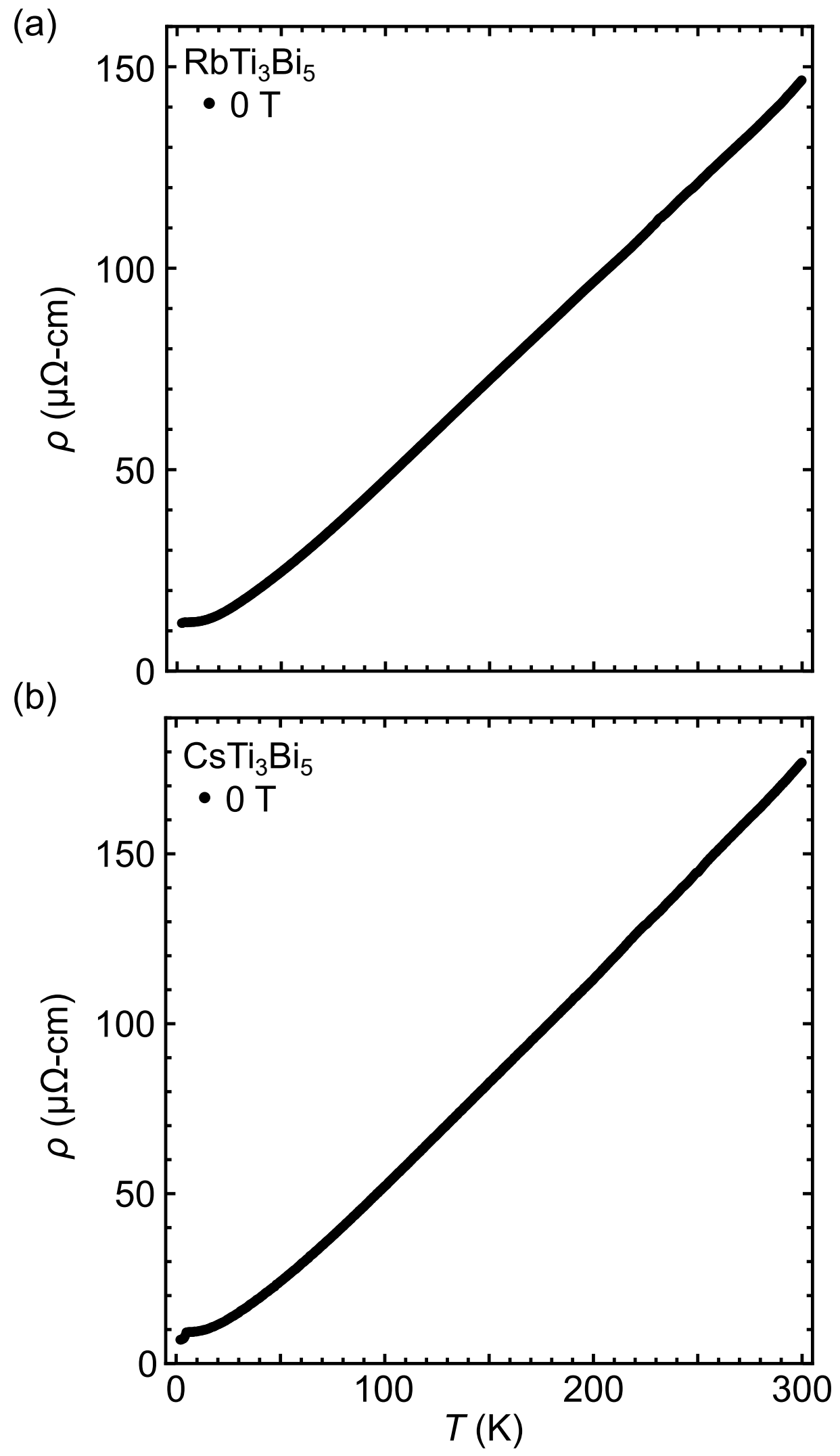

**Figure 5:** In-plane electrical resistivity of $RbTi_3Bi_5$ and $CsTi_3Bi_5$ single crystals at zero field.

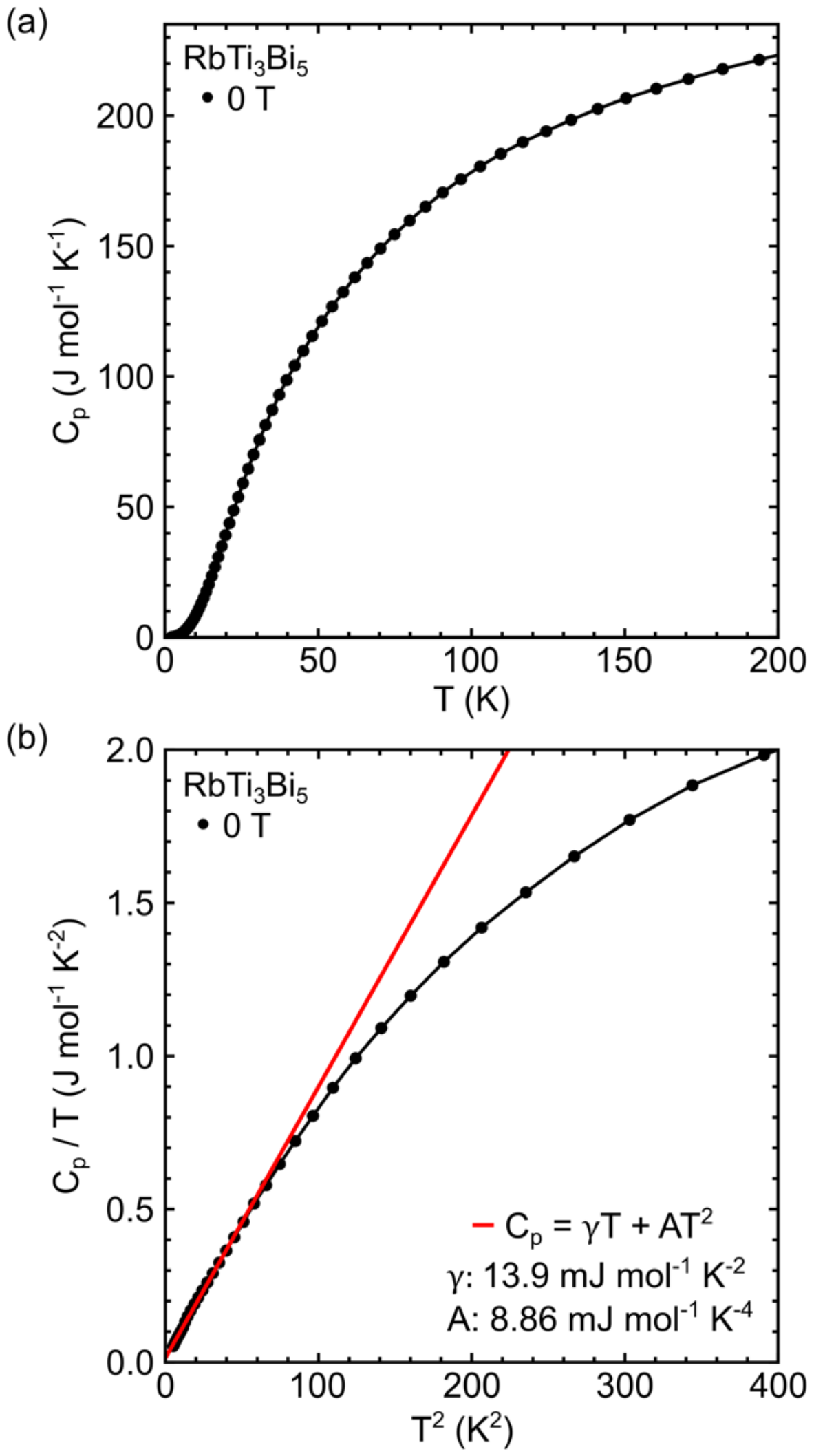

**Figure 6:** Heat capacity of a $RbTi_3Bi_5$ powder compact at zero external field.

Like their sister family $AV_3Sb_5$, the $ATi_3Bi_5$ kagomé metals have exceptionally low resistivities (10–100 μΩ cm) over a wide temperature range, particularly within the *ab* plane. In prior work, we were able to observe quantum oscillations in the magnetoresistance of $CsV_3Sb_5$ (Shubnikov-de Haas effect) [27]. With the natural parallel between the VSb and TiBi systems, we turn to look for quantum oscillations in the $RbTi_3Bi_5$ system. Figure 7a demonstrates the magnetoresistance of a single crystal of $RbTi_3Bi_5$ with current flowing along the *a* axis and the field parallel to the *c* axis. At a 14 T field, the magnetoresistance increases approximately 25% from the zero-field data. Clear signatures of quantum oscillations can be seen in the magnetoresistance.

To extract the oscillatory component of the resistivity, we fit a quartic polynomial background function to the resistivity (red curve, Figure 7a). Subtracting the background function yields the purely oscillatory component (Figure 7b). The Fourier transform of the oscillatory data yields a single major frequency at 200 T (Figure 7c). This is a remarkably simple spectra, considering the relative complexity of the band structure. Observation of many of the higher frequency orbits may be limited due to the lower RRR in the Ti-Bi family and any additional complexities due to air sensitivity.

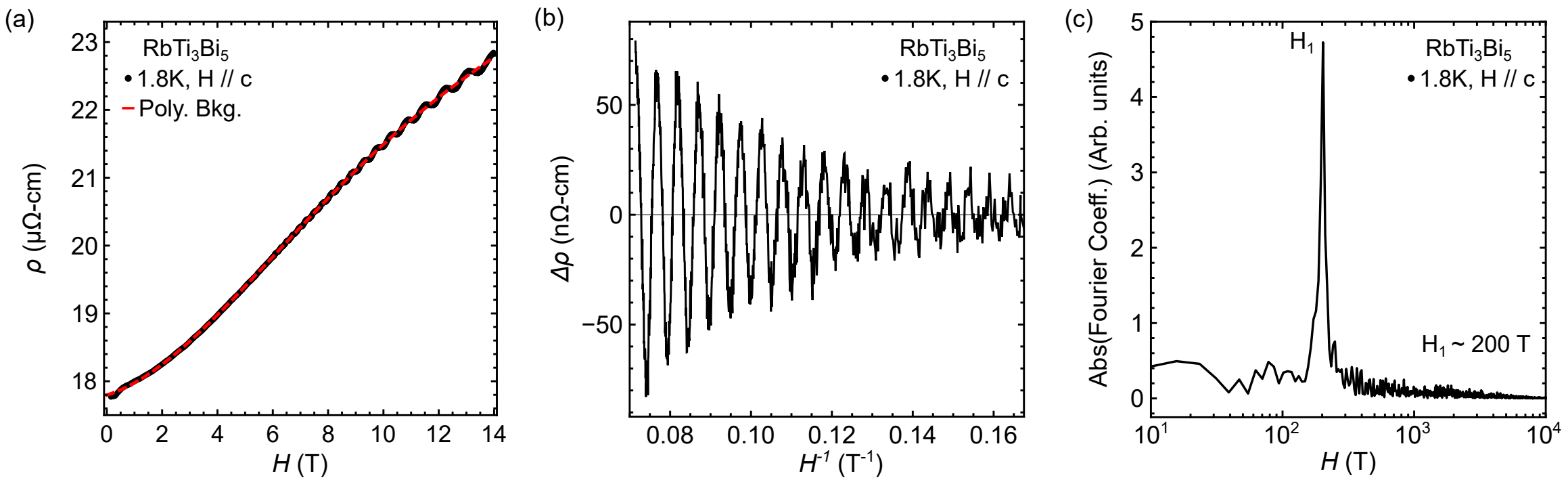

**Figure 7:** Magnetoresistance measurements (a) on $RbTi_3Bi_5$ at 1.8 K show clear signatures of quantum oscillations (Shubnikov-de Haas effect). Utilizing a quartic polynomial to subtract off the background resistance, we isolate the oscillatory component of the magnetoresistance (b) and perform Fourier analysis (c) to identify a dominant frequency ($H_1$ ~ 200 T).

### 2.3 Electronic band structure

The electronic structure of $CsTi_3Bi_5$ was calculated from first principles using the Vienna *Ab initio* Simulation Package (VASP). Figure 8 shows the total and atom-projected density of states (DOS) of $CsTi_3Bi_5$ together with the one of $CsV_3Sb_5$ for comparison.

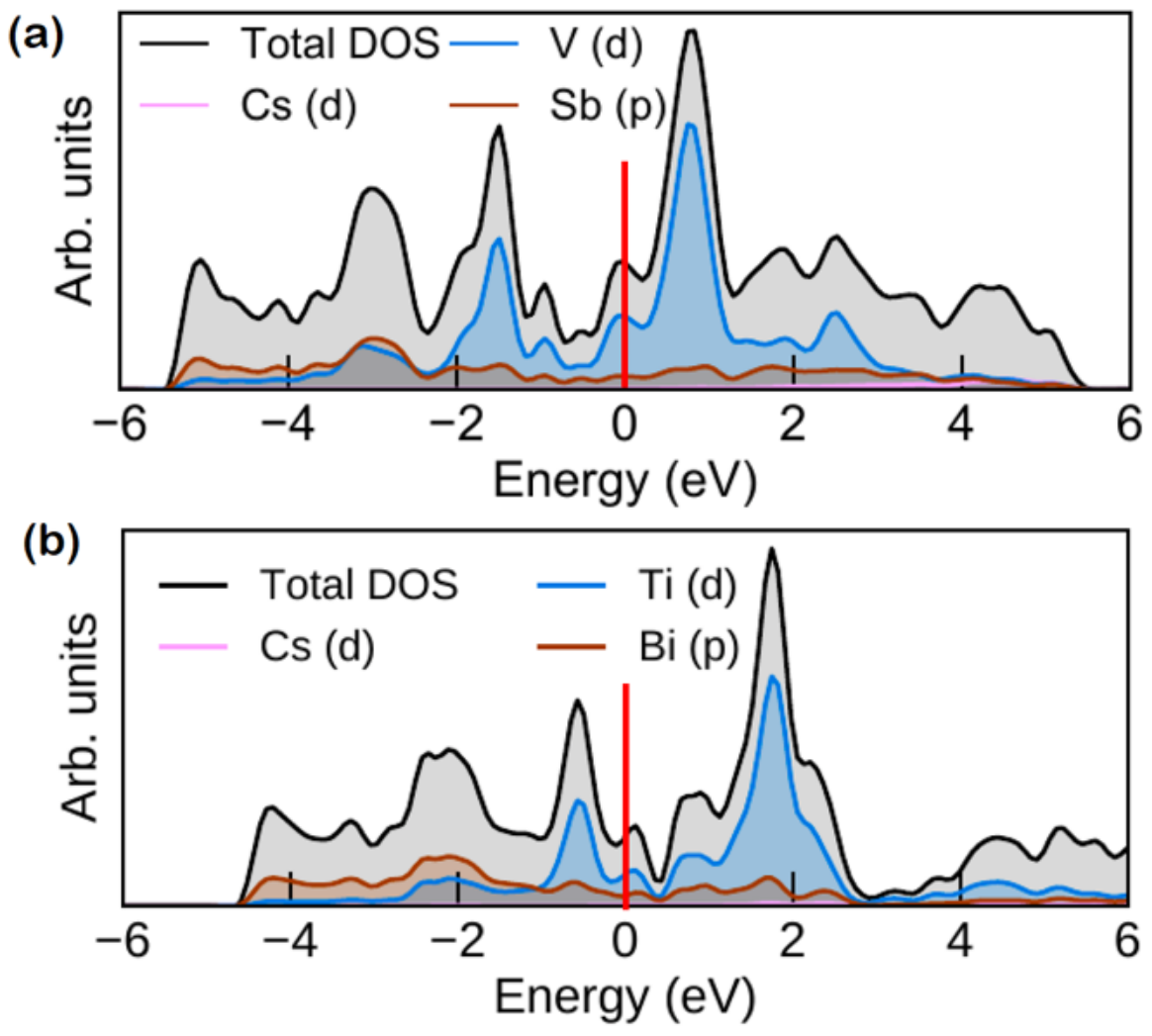

**Figure 8:** Electronic total and projected density of states for $CsV_3Sb_5$ and $CsTi_3Bi_5$

Both DOS patterns are similar and mainly differ by the position of the Fermi level, which is much lower in the case of $CsTi_3Bi_5$. This is as expected, because it has three electrons less per unit cell the thus the band filling is much lower. The effect mainly relates to 3*d* bands (blue shaded areas), which have the largest contributions around the Fermi energy. The band structure of $CsTi_3Bi_5$ is strongly affected by spin-orbit coupling due to the heavy element bismuth. Figure 9 shows the band structure of $CsTi_3Bi_5$ with spin orbit coupling included.

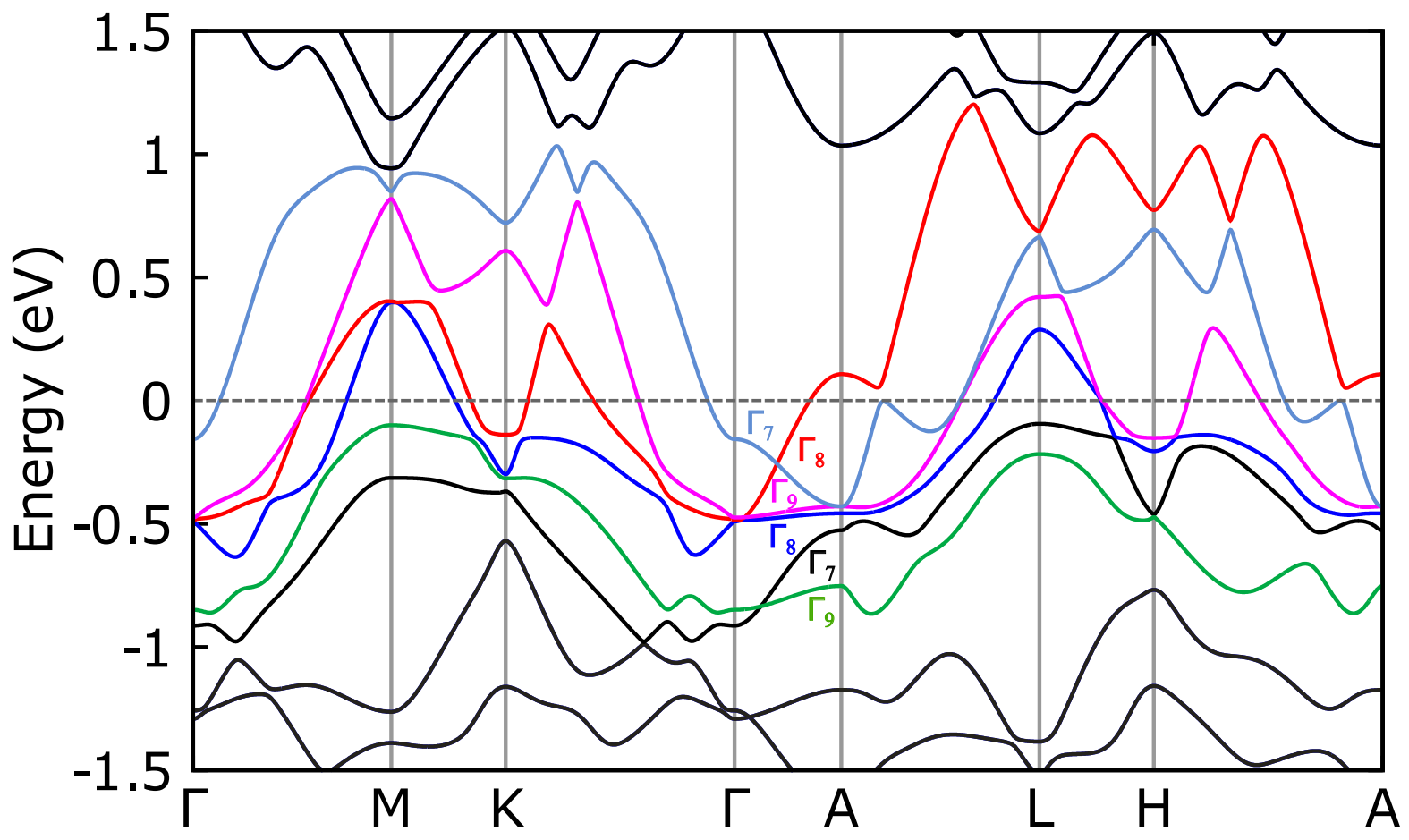

**Figure 9:** Electronic band structure of $CsTi_3Bi_5$ with spin orbit coupling. The symmetry labels at along Γ-A mark symmetry-enforced crossings so that these bands cannot form a gap.

Symmetry analysis of the band structure has characterized the vanadium compounds $AV_3Sb_5$ ($A$ = K, Rb, Cs) as $\mathbb{Z}_2$ topological metals [22, 23]. In these V-based kagomé systems, the presence of continuous gaps between the bands near the Fermi level permitted the $\mathbb{Z}_2$ topological classification via wave function parity product calculations at the time-reversal invariant momentum (TRIM) points in the Brillouin zone [32]. $CsTi_3Bi_5$ likewise possesses time-reversal and inversion symmetry necessary for this $\mathbb{Z}_2$ calculation method, however, analysis of the electronic structure of $CsTi_3Bi_5$ using a higher $k$-point density and considering the band symmetries between the high symmetry points indicate that relevant bands (red, magenta, light blue in Fig. 9) exhibit symmetry-enforced crossings between the $\Gamma_7/\Gamma_8$ bands (red, light blue) and the $\Gamma_8/\Gamma_9$ bands (red, magenta) between the Γ and A points. Given these crossings and the subsequent absence of continuous gaps, the $\mathbb{Z}_2$ topological invariant cannot be calculated using parity products at the TRIM points. Further analysis of the nature of these band crossings is necessary to categorize the electronic structure of $CsTi_3Bi_5$ as topologically trivial or not.

## 3 Experimental Section

Polycrystalline samples of $RbTi_3Bi_5$ and $CsTi_3Bi_5$ were synthesized by heating stoichiometric mixtures of the elements (Rb, ingot, Alfa 99.75%; Cs, liquid, Smart 99.98%; Ti, powder, Alfa 99.5%; Bi, coarse powder, Sigma 99.999%) in alumina crucibles welded in silica ampoules under an atmosphere of purified argon. Both $RbTi_3Bi_5$ and $CsTi_3Bi_5$ are grey crystalline powders interspersed with platelet silver crystals (up to $0.12 \times 0.12 \times 0.04$ mm$^3$) after being sintered at 600°C for 48 h and subsequently cooled at a rate of 20 K h$^{-1}$. Single crystals of $RbTi_3Bi_5$ and $CsTi_3Bi_5$ were synthesized using a self-flux with a method and composition similar to that published previously [27]. Specifically, we created a precursor phase by ball milling elemental Bi shot (Alfa, 99.999%), Ti powder (Alfa, 99.9%), and Cs liquid (Alfa, 99.98%) at a molar ratio of approximately 1:1:6. This precursor was loaded into CoorsTek 2 mL high-density high-purity alumina crucibles and sealed in either: 1) carbon-coated silica ampoules sealed under 0.5 atm argon, or 2) steel tubes sealed under 1 atm argon. Fluxes are heated to 900 °C at a rate of 200 K $h^{-1}$, held at 900°C for 12 h, and cooled to 450–500 °C at a rate of 3 K $h^{-1}$ before molten flux was removed through centrifugation. Crystals form as hexagonal plates with typical dimensions of up to 1 cm × 1cm × 0.25 $mm^3$. They are malleable, readily exfoliated, and mechanically similar to the $AV_3Sb_5$ family. However, unlike $AV_3Sb_5$, crystals are extremely sensitive to air and water, and freshly cleaved surfaces will tarnish irreversibly in the order of 1 minute. Tarnished layers do not protect against additional damage, and degradation will proceed throughout the entire crystal over the course of a few hours.

Due to the air sensitivity of the powders and crystals, all measurement preparation was performed in an argon glovebox with water and oxygen levels <0.5ppm.

The crystal structure of $CsTi_3Bi_5$ was determined by single-crystal X-ray diffraction using a Bruker D8 Quest diffractometer with a fixed χ-goniometer, Mo$K\alpha$ radiation and a Photon 100 detector. Data was acquired using the APEX3 software [33]. The structure was refined using SHELXL [34]. X-ray Powder diffraction data was measured on a STOE Stadi-P diffractometer using a Mo$K\alpha$ source and a Dectris Mythen 1K strip detector. Powders were diluted in a mass ratio of 1:5 with ground amorphous glass and sealed in glass capillaries ($d$ = 0.1 mm). The structure of $CsTi_3Bi_5$ were used as starting parameters for Rietveld fits with TOPAS V4.1 [35].

Further details of the crystal structure investigation may be obtained from Fachinformationszentrum Karlsruhe, 76344 Eggenstein-Leopoldshafen, Germany (fax: +49-7247-808-666; e-mail: crysdata@fiz-karlsruhe.de, http://www.fiz-infomationsdienste.de/en/DB/icsd/depot_anforderung.html) on quoting the deposition numbers CSD 2208937 ($CsTi_3Bi_5$) and CSD 2208938 ($RbTi_3Bi_8$).

Magnetization measurements were performed on a Quantum Design MPMS3 using vibrating sample magnetometry (VSM). Superconductivity tests were performed under a 5 Oe applied field, while general screening for other magnetic properties was performed at 1 T. Powder samples were loaded in polypropylene capsules, sealed with a thin layer of PARAFILM®. Crystals were mounted onto quartz paddles with small quantities of GE varnish. Resistivity and quantum oscillation measurements were performed on a Quantum Design Dynacool 14T system using the electrical transport option (ETO) with a probe current of 5 mA and frequency of ~100 Hz. Crystals were cleaved, exfoliated, and mounted to the resistivity stage within the glovebox. Contacts were made using silver paint (DuPont cp4929N-100) and gold wire (Alfa, 0.05 mm premion 99.995%). Heat capacity measurements were performed on a Quantum Design Dynacool 9 T system using the heat capacity option. Sections of a sintered powder compact were adhered to the sample stage using a small quantity of Apezion N grease.

The electronic structure of $CsTi_3Bi_5$ was calculated from first principles using the Vienna *Ab initio* Simulation Package (VASP, version 5.4.4) [36–38] based on density functional theory and plane wave basis sets. These calculations employed the PBE functional with D3 corrections to account for the van-der-Waals forces along the *c* axis in $CsTi_3Bi_5$ [39]. Projector-augmented wave (PAW) pseudopotentials [40] were selected based on the VASP 5.4 recommendations. The plane wave energy cutoff was set to 520 eV and a 23 × 23 × 11 Γ-centered *k*-mesh was automatically generated within VASP using the Monkhorst-Pack method [41]. The structure was relaxed with a force cutoff of $10^{-3}$ eV Å$^{-1}$. All calculations were performed with spin-orbit coupling and an energy convergence better than $10^{-7}$ eV. The *k*-point path for the band structure (Figure 9) was generated using SUMO-KGEN [42] and contains 1899 *k*-points.

**Acknowledgements:** This work was financially supported by the German Research Foundation (DFG) and the Bavaria California Technology Center (BaCaTeC, Grant 7 [2021-2]). The authors acknowledge the computational and data resources provided by the Leibniz Supercomputing Centre (www.lrz.de). We gratefully thank Maia Garcia-Vergniory and Iñigo Robredo for useful advice. The work at UC Santa Barbara was supported by the National Science Foundation (NSF) through Enabling Quantum Leap: Convergent Accelerated Discovery Foundries for Quantum Materials Science, Engineering and Information (Q-AMASE-i): Quantum Foundry at UC Santa Barbara (Grant No. DMR-1906325). The use of shared facilities of the NSF Materials Research Science and Engineering Center at UC Santa Barbara Grant No. DMR-1720256, a member of the Materials Research Facilities Network is acknowledged. S.D.W. and B.R.O. acknowledge support by the US Department of Energy Office of Basic Energy Sciences, Division of Materials Science and Engineering under award DE-SC0020305

Note added: During the preparation of this manuscript, we recognized unpublished preprints about $CsTi_3Bi_5$ (https://arxiv.org/abs/2209.03840 and https://arxiv.org/abs/2202.05588).

**References**

1. Grey I. E. *Mineral. Mag.* 2020, *84*, 640–652.
2. Stein F., Leineweber A. *J. Mater. Sci.* 2021, *56*, 5321–5427.
3. Syôzi I. *Prog. Theor. Phys.* 1951, *6*, 306-308.
4. Anderson P. W. *Mater. Res. Bull.* 1973, *8*, 153–160.
5. Sachdev S. *Phys. Rev. B* 1992, *45*, 12377–12396.
6. Anderson P. W. *Science* 1987, *235*, 1196–1198.
7. Hiroi Z., Ishikawa H., Yoshida H., Yamaura J.-i., Okamoto Y. *Inorg. Chem.* 2019, *58*, 11949–11960.
8. Gingras M. J. P., Stager C. V., Raju N. P., Gaulin B. D., Greedan J. E. *Phys. Rev. Lett.* 1997, *78*, 947–950.
9. Shores M. P., Nytko E. A., Bartlett B. M., Nocera D. G. *J. Am. Chem. Soc.* 2005, *127*, 13462–13463.
10. Helton J. S., Matan K., Shores M. P., Nytko E. A., Bartlett B. M., Yoshida Y., Takano Y., Suslov A., Qiu Y., Chung J. H., Nocera D. G., Lee Y. S. *Phys. Rev. Lett.* 2007, *98*, 107204 (4 pages).
11. Okamoto Y., Nohara M., Aruga-Katori H., Takagi H. *Phys. Rev. Lett.* 2007, *99*, 137207 (4 pages).
12. Mendels P., Bert F., de Vries M. A., Olariu A., Harrison A., Duc F., Trombe J. C., Lord J. S., Amato A., Baines C. *Phys. Rev. Lett.* 2007, *98*, 077204 (4 pages).
13. Depenbrock S., McCulloch I. P., Schollwöck U. *Phys. Rev. Lett.* 2012, *109*, 067201 (6 pages).
14. Yan S., Huse D. A., White S. R. *Science* 2011, *332*, 1173–1176.
15. Broholm C., Cava R. J., Kivelson S. A., Nocera D. G., Norman M. R., Senthil T. *Science* 2020, *367*, eaay0668 (9 pages).
16. Guo H. M., Franz M. *Phys. Rev. B* 2009, *80*, 113102 (4 pages).
17. O'Brien A., Pollmann F., Fulde P. *Phys. Rev. B* 2010, *81*, 235115 (10 pages).
18. Wang W.-S., Li Z.-Z., Xiang Y.-Y., Wang Q.-H. *Phys. Rev. B* 2013, *87*, 115135 (8 pages).

19. Ortiz B. R., Gomes L. C., Morey J. R., Winiarski M., Bordelon M., Mangum J. S., Oswald L. W. H., Rodriguez-Rivera J. A., Neilson J. R., Wilson S. D., Ertekin E., McQueen T. M., Toberer E. S. *Phys. Rev. Mater.* 2019, *3*, 094407 (9 pages).

20. Jiang Y. X., Yin J. X., Denner M. M., Shumiya N., Ortiz B. R., Xu G., Guguchia Z., He J. Y., Hossain M. S., Liu X. X., Ruff J., Kautzsch L., Zhang S. T. S., Chang G. Q., Belopolski I., Zhang Q., Cochran T. A., Multer D., Litskevich M., Cheng Z. J., Yang X. P., Wang Z. Q., Thomale R., Neupert T., Wilson S. D., Hasan M. Z. *Nat. Mater.* 2021, *20*, 1353–1357.

21. Ortiz B. R., Sarte P. M., Kenney E. M., Graf M. J., Teicher S. M. L., Seshadri R., Wilson S. D. *Phys. Rev. Mater.* 2021, *5*, 034801 (7 pages).

22. Ortiz B. R., Teicher S. M. L., Hu Y., Zuo J. L., Sarte P. M., Schueller E. C., Abeykoon A. M. M., Krogstad M. J., Rosenkranz S., Osborn R., Seshadri R., Balents L., He J. F., Wilson S. D. *Phys. Rev. Lett.* 2020, *125*, 247002 (6 pages).

23. Yang S. Y., Wang Y. J., Ortiz B. R., Liu D. F., Gayles J., Derunova E., Gonzalez-Hernandez R., Smejkal L., Chen Y. L., Parkin S. S. P., Wilson S. D., Toberer E. S., McQueen T., Ali M. N. *Sci. Adv.* 2020, *6*, eabb6003 (7 pages).

24. Oey Y. M., Kaboudvand F., Ortiz B. R., Seshadri R., Wilson S. D. *Phys. Rev. Mater.* 2022, *6*, 074802 (7 pages).

25. Oey Y. M., Ortiz B. R., Kaboudvand F., Frassineti J., Garcia E., Cong R., Sanna S., Mitrović V. F., Seshadri R., Wilson S. D. *Phys. Rev. Mater.* 2022, *6*, L041801 (6 pages).

26. Jiang Y., Yu Z., Wang Y., Lu T., Meng S., Jiang K., Liu M. Chin. Phys. Lett. 2022, 39, 047402-1-047402-7.

27. Ortiz B. R., Teicher S. M. L., Kautzsch L., Sarte P. M., Ratcliff N., Harter J., Ruff J. P. C., Seshadri R., Wilson S. D. *Phys. Rev. X* 2021, *11*, 041030 (14 pages).

28. Hoffmann R.-D., Pöttgen R. *Z. Kristallogr.* 2001, *216*, 127–145.

29. Jovanovic M., Schoop L. M. *J. Am. Chem. Soc.* 2022, *144*, 10978–10991.

30. Li H., Ikeda M., Suzuki A., Taguchi T., Zhang Y., Goto H., Eguchi R., Liao Y.-F., Ishii H., Kubozono Y. *Phys. Chem. Chem. Phys.* 2022, *24*, 7185–7194.

31. Roberts B. W. *J. Phys. Chem. Ref. Data* 1976, *5*, 581–821.

32. Fu L., Kane C. L. *Phys. Rev. B* 2007, *76*, 045302 (17 pages).

33. APEX3, Data Reduction and Frame Integration Program for the CCD Area-Detector System, Bruker AXS Inc.: Madison, Wisconsin (USA), 2016.

34. Sheldrick G. M. *Acta Crystallogr.* 2008, *A64*, 112–122.

35. Coelho A. TOPAS-Academic (version 4.1), Coelho Software: Brisbane (Australia), 2007.

36. Kresse G., Furthmüller J. *Comp. Mater. Sci.* 1996, *6*, 15–50.

37. Kresse G., Furthmüller J. *Phys. Rev. B* 1996, *54*, 11169–11186.

38. Kresse G., Hafner J. *Phys. Rev. B* 1993, *47*, 558–561.

39. Grimme S., Antony J., Ehrlich S., Krieg H. *J. Chem. Phys.* 2010, *132*, 154104 (19 pages).

40. Kresse G., Joubert D. *Phys. Rev. B* 1999, *59*, 1758–1775.

41. Monkhorst H. J., Pack J. D. *Phys. Rev. B* 1976, *13*, 5188–5192.

42. Ganose A. M., Jackson A. J., Scanlon D. O. *Journal of Open Source Software* 2018, *3*, 717 (3 pages).